\documentclass[aps,prd,onecolumn,groupedaddress,showpacs,nofootinbib,amssymb
]{revtex4}
\usepackage[dvips]{graphicx}
\usepackage{amssymb}
\usepackage{amsmath}
\usepackage{graphicx,color}
\usepackage{amsfonts}
\usepackage{bm}
\begin{document}

\tolerance=5000

\title{Did the Universe Experienced a Pressure non-Crushing Type Cosmological Singularity in the Recent Past?}
\author{S.D. Odintsov,$^{1,2,3}$\,\thanks{odintsov@ieec.uab.es}
V.K. Oikonomou,$^{4,5}$\,\thanks{v.k.oikonomou1979@gmail.com}}
\affiliation{$^{1)}$ ICREA, Passeig Luis Companys, 23, 08010 Barcelona, Spain\\
$^{2)}$ Institute of Space Sciences (ICE,CSIC) C. Can Magrans s/n,
08193 Barcelona, Spain\\
$^{3)}$ Institute of Space Sciences of Catalonia (IEEC),
Barcelona, Spain\\
$^{4)}$ Department of Physics, Aristotle University of Thessaloniki, Thessaloniki 54124, Greece\\
$^{5)}$Laboratory for Theoretical Cosmology, International Center
of Gravity and Cosmos, Tomsk State University of Control Systems
and Radioelectronics  (TUSUR), 634050 Tomsk, Russia }

\begin{abstract}
In the recent literature it has been shown that the $H_0$ tension
may be eliminated if an abrupt physics transition changed the
Cepheid parameters in the near past of the Universe, nearly
$70-150\,$Myrs ago. In this letter we stress the possibility that
this abrupt transition was caused by the smooth passage of our
Universe through a pressure finite-time cosmological singularity.
Being a non-crushing type singularity the pressure singularity can
leave its imprints in the Universe, since it occurs globally and
literally everywhere. We discuss how this scenario could easily be
realized by $F(R)$ gravity, with the strong energy conditions
being satisfied without the need for a scalar field or specific
matter fluids. We also stress the fact that the pressure
singularity can affect the effective gravitational constant of
$F(R)$ gravity. Moreover, we stress the fact that pressure
singularities can disrupt the trajectories of bound objects in the
Universe, which is also pointed out in the literature, even in the
context of general relativity. We also show numerically in a
general relativistic framework that elliptic trajectories are
distorted and changed to different elliptic trajectories when the
Universe passes through the pressure singularity. Such a
disruption of the trajectories could have tidal effects on the
surface of the earth, for example on sea waters  and oceans,
regarding the distortion of moon's elliptic trajectory.
Accordingly, the distortion of earth's trajectory around the sun
could have affected climatologically earth $70-150\,$Myrs ago.
\end{abstract}

%PACS numbers: 04.50.Kd, 95.36.+x, 98.80.-k, 98.80.Cq
%\pacs{04.50.Kd, 95.36.+x, 98.80.-k, 98.80.Cq,11.25.-w}

\maketitle

\section{Introduction}

The $H_0$ tension \cite{Perivolaropoulos:2021jda,Verde:2019ivm} is
an undoubtable fact according to the latest observations. Such
discrepancy between the Cosmic Microwave Background measurements
and the low-redshift sources on the present day $H_0$ could be a
calibration argument, as was pointed out in Refs.
\cite{Mortsell:2021nzg,Perivolaropoulos:2021bds}. As was pointed
out in Refs. \cite{Perivolaropoulos:2021bds,Marra:2021fvf}, the
$H_0$ tension is eliminated if an abrupt transition in the Cepheid
parameters have occurred before $70-150\, $Myrs. Such a change
could have occurred if the physics of the Universe globally
changed around $70-150\, $Myrs ago, and this could also have
affected the strength of the gravitational force during the same
period of time. In this letter we aim to discuss the possibility
that such a physics change may have been caused by the smooth
passage of the Universe through a pressure singularity. Pressure
singularities are mild non-crushing type singularities, which also
respect the strong energy conditions. It is known that such
singularities cause deformations in the trajectories of bound
systems \cite{Perivolaropoulos:2016nhp}. Thus, if the Universe
passed through a pressure singularity before $70-150\, $Myrs, it
is possible that the trajectories of the bound planets in the
solar system could also have been affected. In addition the orbit
of the moon around earth could have been deformed too. Thus it is
possible that these changes in the orbits could have generated
tidal effects on the surface of the earth, like elimination of
movement of sea waters or oceans (moon-earth system), or even
climatological changes on earth (earth-sun system). We discuss how
a pressure singularity could have been generated by an $F(R)$
gravity, how this pressure singularity could have affected the
effective gravitational constant. Also, following the analysis and
presentation of \cite{Perivolaropoulos:2016nhp}, we explicitly
show in a general relativistic (GR) framework how elliptic
trajectories are deformed when the Universe passes through a
pressure singularity.

For this work we shall assume that the geometric background is
described by a flat Friedmann-Robertson-Walker (FRW) metric, with
its line element being,
\begin{equation}\label{frw}
ds^2 = - dt^2 + a(t)^2 \sum_{i=1,2,3} \left(dx^i\right)^2\, ,
\end{equation}
where $a(t)$ is the scale factor, while the Ricci scalar for the
FRW metric is
\begin{equation}\label{ricciscalaranalytic}
R=6\left (\dot{H}+2H^2 \right )\, ,
\end{equation}
where $H=\frac{\dot{a}}{a}$, denotes as usual the Hubble rate.

\section{The Perspective of a Pressure Singularity in the Near Past and the $F(R)$ Gravity Picture}

Following the argument of Ref. \cite{Perivolaropoulos:2016nhp}, it
is probable that the $H_0$ tension might be an artifact of
calibration. As was shown and claimed in
\cite{Perivolaropoulos:2016nhp}, an abrupt change of physics
before $70-150\,$Myrs may, can directly affect the Cepheid
parameters and can eliminate the Hubble tension. Following this
line of research, we shall discuss the possibility that this
abrupt change in the physics of the Universe, is caused by a
global finite-time cosmological singularity of sudden type. These
singularities are known to be geodetically complete, and also the
strong energy conditions are respected when the Universe passes
through these smooth singularities. In the following we shall
assume that the spacetime metric is a flat FRW metric. Let us
recall the classification of finite-time cosmological
singularities, which are classified as follows
\cite{Nojiri:2005sx} if the time instance of the singularity is at
a finite-time $t_s$:
\begin{itemize}
\item Type I (``Big Rip'') : This is a crushing type singularity,
in which case, as the singularity at $t \to t_s$, the scale factor
$a(t)$, the total effective pressure $p_\mathrm{eff}$ and energy
density $\rho_\mathrm{eff}$, diverge, so effectively $a \to
\infty$ , $\rho_\mathrm{eff} \to \infty$, and
$\left|p_\mathrm{eff}\right| \to \infty$ \cite{bigrip}. \item Type
II (``sudden''): This a pressure singularity, firstly studied in
Refs. \cite{barrowsudden}, see also \cite{barrowsudden1}, for
which as $t \to t_s$, $a \to a_s$, $\rho_\mathrm{eff} \to \rho_s$,
$\left|p_\mathrm{eff}\right| \to \infty$. This is the type of
cosmological singularity for which we shall be interested for in
this paper. \item Type III : In this case as $t \to t_s$, we have
$a \to a_s$, $\rho_\mathrm{eff} \to \infty$,
$\left|p_\mathrm{eff}\right| \to \infty$. \item Type IV : This
type of cosmological singularity is the mildest singularity
studied thoroughly in
\cite{Nojiri:2005sx,Nojiri:2004pf,Barrow:2015ora,Nojiri:2015fra,Nojiri:2015fia,Odintsov:2015zza,Oikonomou:2015qha,Oikonomou:2015qfh},
for which as  $t\to t_s$, we have $a \to a_s$, $\rho_\mathrm{eff}
\to \rho_s$, $\left|p_\mathrm{eff}\right| \to p_s$, while
$\frac{\mathrm{d}^nH}{\mathrm{d}t^n}$, for $n\geq 2$ diverge.
\end{itemize}
We shall be interested for pressure singularities, or sudden
singularities \cite{barrowsudden}. As we already mentioned, for
these singularities, the total effective energy density, the scale
factor and its first time derivative, and the Hubble rate, are
finite, as the singularity is approached. However, the first
derivative of the Hubble rate, the pressure and the second
derivative of the scale factor are divergent as the singularity is
approached. In order to have a more quantitative idea on the
evolution of the Universe near a pressure singularity, let us
assume that the scale factor has the following form,
\begin{equation}\label{scalefactorini}
a(t)\simeq g(t)(t-t_s)^{\alpha}+f(t)\, ,
\end{equation}
where $g(t)$ and $f(t)$ and all their higher order derivatives
with respect to the cosmic time are smooth functions of the cosmic
time. For consistency in order not to have complex values before
and after the singularity, let us assume that $a=\frac{2m}{2n+1}$.
According to the values of the exponent $\alpha$, the following
types of singularities may occur,
\begin{itemize}
\item For $\alpha <0$ the Universe develops a Type I singularity.
\item For $0<\alpha<1$  the Universe develops a Type III
singularity. \item For $1<\alpha<2$  the Universe develops a Type
II singularity. \item For $2<\alpha$  the Universe develops a Type
IV singularity.
\end{itemize}
Thus the pressure singularity occurs for $1<\alpha<2$, since the
second derivative of the scale factor near the singularity is
$\ddot{a}\sim (\alpha -1) \alpha  g(t) (t-t_s)^{\alpha -2}$, and
the same applies for the derivative of the Hubble rate, which near
the singularity becomes at leading order $\dot{H}\sim
\frac{(\alpha -1) \alpha  (t-t_s)^{\alpha -2}}{f(t)}$.

For pressure singularities, the only physical quantity that
diverges is the pressure, thus the energy density remains finite.
The singularity occurs in a global way, so at $t=t_s$ which is a
spacelike hypersurface, the pressure diverges. However, a pressure
singularity does not lead to geodesics incompleteness, since for
all causal geodesics, at $t=t_s$, the following integral is finite
\cite{Fernandez-Jambrina:2004yjt},
\begin{equation}\label{highercurvsc}
\int_0^{\tau}dt R^{i}_{0j0}(t)\, .
\end{equation}
Regarding the strong energy conditions, these are not violated at
a pressure singularity, but we have
$\rho_\mathrm{eff}+3p_\mathrm{eff}>0$ and $\rho_\mathrm{eff}>0$.
This condition can be realized by many GR cosmological frameworks,
like from scalar fields and cosmological fluids. In this article,
for the moment, we shall assume that the scale factor $a(t)$ of
Eq. (\ref{scalefactorini}) is realized by $F(R)$ gravity. In fact
it is much more feasible to realize the scale factor
(\ref{scalefactorini}) and at the same time to respect the strong
energy conditions $\rho_\mathrm{eff}+3p_\mathrm{eff}>0$ in $F(R)$
gravity. The reason is that in the case of $F(R)$ gravity, the
energy momentum tensor receives a geometric contribution, but let
us give some details on this. Consider the $F(R)$ gravity action
in the presence of matter fluids,
\begin{equation}\label{action}
\mathcal{S}=\frac{1}{2\kappa^2}\int
\mathrm{d}^4x\sqrt{-g}F(R)+S_m(g_{\mu \nu},\Psi_m),
\end{equation}
where $\kappa^2=8\pi G$ and also $S_m$ denotes the matter fluids
present. Note that future singularities in $F(R)$ gravity were
thoroughly studied in \cite{Nojiri:2006ri}. In the metric
formalism, the field equations read,
\begin{align}\label{modifiedeinsteineqns}
R_{\mu \nu}-\frac{1}{2}Rg_{\mu
\nu}=\frac{\kappa^2}{F_R(R)}\Big{(}T_{\mu
\nu}+\frac{1}{\kappa^2}\Big{(}\frac{F(R)-RF_R(R)}{2}g_{\mu
\nu}+\nabla_{\mu}\nabla_{\nu}F_R(R)-g_{\mu \nu}\square
F_R(R)\Big{)}\Big{)}\, .
\end{align}
where, $F_R(R)=\partial F(R)/\partial R$ and $T_{\mu \nu}$ stands
for the energy momentum tensor of the perfect matter fluids that
are present. We can write the $F(R)$ gravity field equations in
the Einstein-Hilbert form, ,
\begin{equation}\label{geometridde}
R_{\mu \nu}-\frac{1}{2}Rg_{\mu \nu}=T_{\mu \nu}^{m}+T_{\mu
\nu}^{curv}\, ,
\end{equation}
where $T_{\mu \nu}^{m}$ is equal to,
\begin{equation}\label{tmnmat}
T_{\mu \nu}^{m}=\frac{1}{\kappa}\frac{T_{\mu \nu}}{F_R(R)}\, ,
\end{equation}
and originating from ordinary matter perfect fluids, while $T_{\mu
\nu}^{curv}$ is equal to,
\begin{equation}\label{tmnmatcurv}
T_{\mu
\nu}^{curv}=\frac{1}{\kappa}\Big{(}\frac{F(R)-RF_R(R)}{2}g_{\mu
\nu}+F_R(R)^{;\mu \nu}(g_{\alpha \mu}g_{\beta \nu }-g_{\alpha
\beta}g_{\mu \nu })\Big{)}\, ,
\end{equation}
The energy momentum tensor (\ref{tmnmatcurv}) is thus the
contribution of $F(R)$ gravity to the total energy momentum
tensor. In order to further see this for a flat FRW metric
(\ref{frw}), we consider the $f(R)$ gravity field equations for a
FRW metric, which are
\begin{align}\label{eqnsofmkotion}
& 3 H^2F_R=\frac{RF_R-F}{2}-3H\dot{F}_R+\kappa^2\left(
\rho_r+\rho_m\right)\, ,\\ \notag &
-2\dot{H}F=\ddot{F}_R-H\dot{F}_R +\frac{4\kappa^2}{3}\rho_r\, ,
\end{align}
where $\rho_m$ and $\rho_r$ denote the total energy density and
pressure of the dark matter and radiation fluids present, and
accordingly $p_r$ is the radiation fluid pressure. Rewriting the
field equations in the Einstein-Hilbert form, we have,
\begin{align}\label{flat}
& 3H^2=\kappa^2\rho_{tot}\, ,\\ \notag &
-2\dot{H}=\kappa^2(\rho_{tot}+P_{tot})\, ,
\end{align}
where $\rho_{tot}=\rho_{m}+\rho_{G}+\rho_r$ stands for the total
energy density, $\rho_m$ is dark matter energy density, and
$\rho_r$ is the radiation energy density. Finally, $\rho_{G}$ is
the energy density contribution of the $F(R)$ gravity, which is
defined as,
\begin{equation}\label{degeometricfluid}
\kappa^2\rho_{G}=\frac{F_R R-F}{2}+3H^2(1-F_R)-3H\dot{F}_R\, ,
\end{equation}
and its pressure is,
\begin{equation}\label{pressuregeometry}
\kappa^2P_{G}=\ddot{F}_R-H\dot{F}_R+2\dot{H}(F_R-1)-\kappa^2\rho_{G}\,
.
\end{equation}
Thus the strong energy conditions near a sudden pressure
singularity can be easily satisfied by an $F(R)$ gravity, and at
the same time the scale factor (\ref{scalefactorini}) can be
realized by the same $F(R)$ gravity in a consistent and easy way.

Now let us turn to the physical implications of a pressure
singularity in a FRW Universe described by $F(R)$ gravity.
Firstly, the divergence of the second time derivative of the scale
factor implies a sudden change in the acceleration of the Universe
at the singularity. The Universe may thus superaccelerate or
superdecelerate for a small amount of time before and after the
singularity, depending on the specific form of the scale factor.
This change is an abrupt change in the physics, and in principle
can be aligned with the arguments of
\cite{Perivolaropoulos:2016nhp} for abrupt changes in the physics
before $70-150\,$Myrs.

More importantly, a pressure singularity may directly affect the
effective gravitational constant in $F(R)$ gravity. In the
sub-horizon approximation where the wavenumbers of the modes
satisfy,
\begin{equation}\label{subhorapprx}
\frac{k^2}{a^2}\gg H^2\, ,
\end{equation}
the matter density perturbations quantified by the parameter
$\delta =\frac{\delta \varepsilon_m}{\varepsilon_m}$, satisfy,
\begin{equation}\label{matterperturb}
\ddot{\delta}+2H\dot{\delta}-4\pi G_{eff}(a,k)\varepsilon_m\delta
=0\, ,
\end{equation}
where $G_{eff}(a,k)$ stands for the effective gravitational
constant of gravity $F(R)$ theory which is defined as
\cite{Bamba:2012qi},
\begin{equation}\label{geff}
G_{eff}(a,k)=\frac{G}{F_R(R)}\Big{[}1+\frac{\frac{k^2}{a^2}\frac{F_{RR}(R)}{F_R(R)}}{1+3\frac{k^2}{a^2}\frac{F_{RR}(R)}{F_R(R)}}
\Big{]}\, ,
\end{equation}
where $G$ is Newton's gravitational constant. Also $\varepsilon_m$
in the definition of the matter density perturbation parameter
$\delta$, is the total energy density of the dark matter perfect
fluid. Obviously, the presence of a pressure singularity in the
evolution of the Universe will directly affect the terms in the
effective gravitational constant (\ref{geff}) which are related
with the $F(R)$ function and its derivatives with respect to the
Ricci scalar. This is due to the fact that the Ricci scalar in a
flat FRW spacetime is of the form $R=12H^2+6\dot{H}$, and
$\dot{H}$ is singular at the time instance that the pressure
singularity occurs. Thus, this abrupt change in the effective
gravitational constant can also affect the Cepheid parameters,
thus a pressure singularity in the near past may be responsible
for the abrupt change of physics in the near past. In fact, as was
also pointed out in Ref. \cite{Marra:2021fvf}, a change in the
effective gravitational constant at small redshifts, may remedy
the $H_0$ tension.

So far in this article, we discussed the implications of a
pressure singularity in the evolution of the Universe in the near
past in a theoretical way. In this way, it is not so apparent how
a pressure singularity in the near past evolution of the Universe,
may actually be identified as a mean to remedy the $H_0$ tension.
It is thus vital to have a way to identify such a scenario in an
direct or indirect experimental way. Such a proposal may be
realized by considering the changes on the surface of earth for
cosmic times in the range $70-150\, $Myrs. The reason is simple,
if a sudden pressure singularity occurred in the near past in our
Universe, then it would affect the tidal forces of gravitationally
bound objects and it would certainly disturb their orbits. Such a
scenario was considered thoroughly in Ref.
\cite{Perivolaropoulos:2016nhp}, and it was shown that a sudden
pressure singularity can indeed disturb the orbits of bound
objects. If this scenario indeed took place in the Universe, it
would also affect the solar system, and of course it would have
direct implications on the orbit of the moon around earth.
\begin{figure}
\centering
\includegraphics[width=18pc]{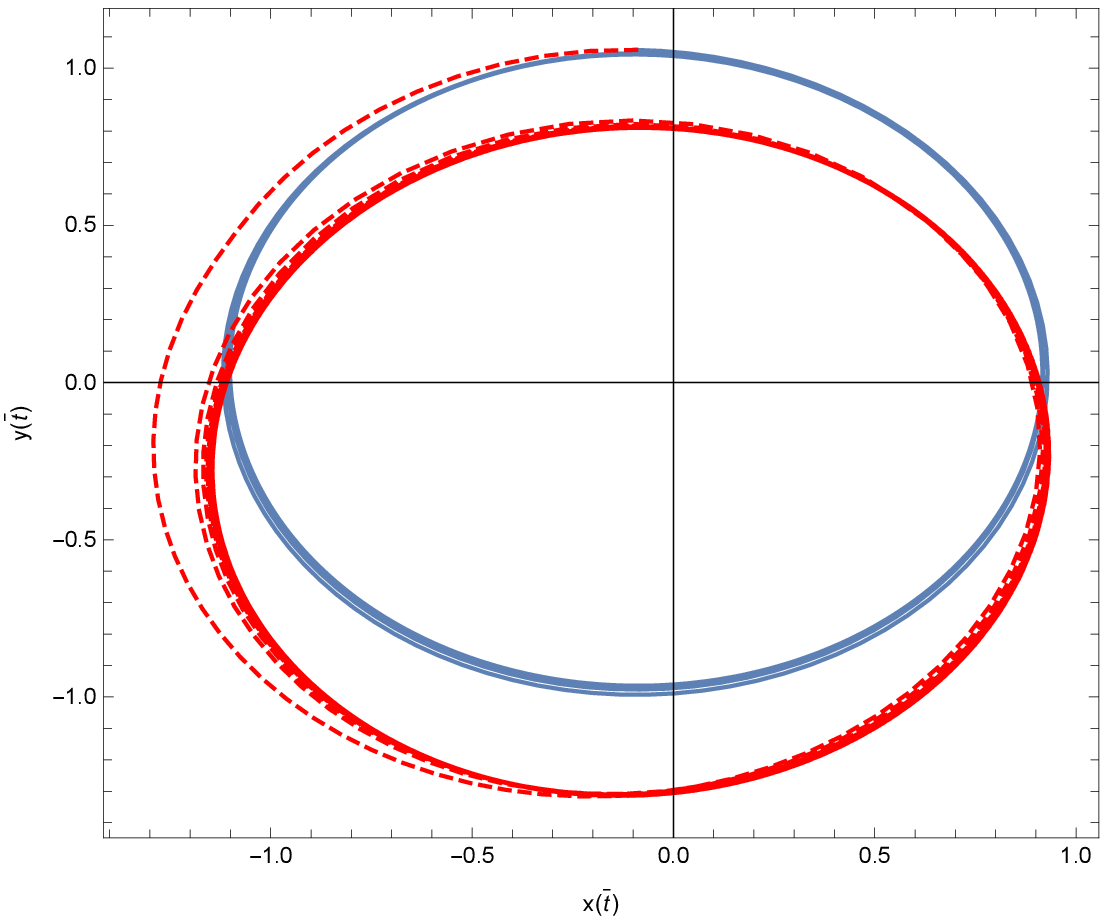}
\includegraphics[width=15pc]{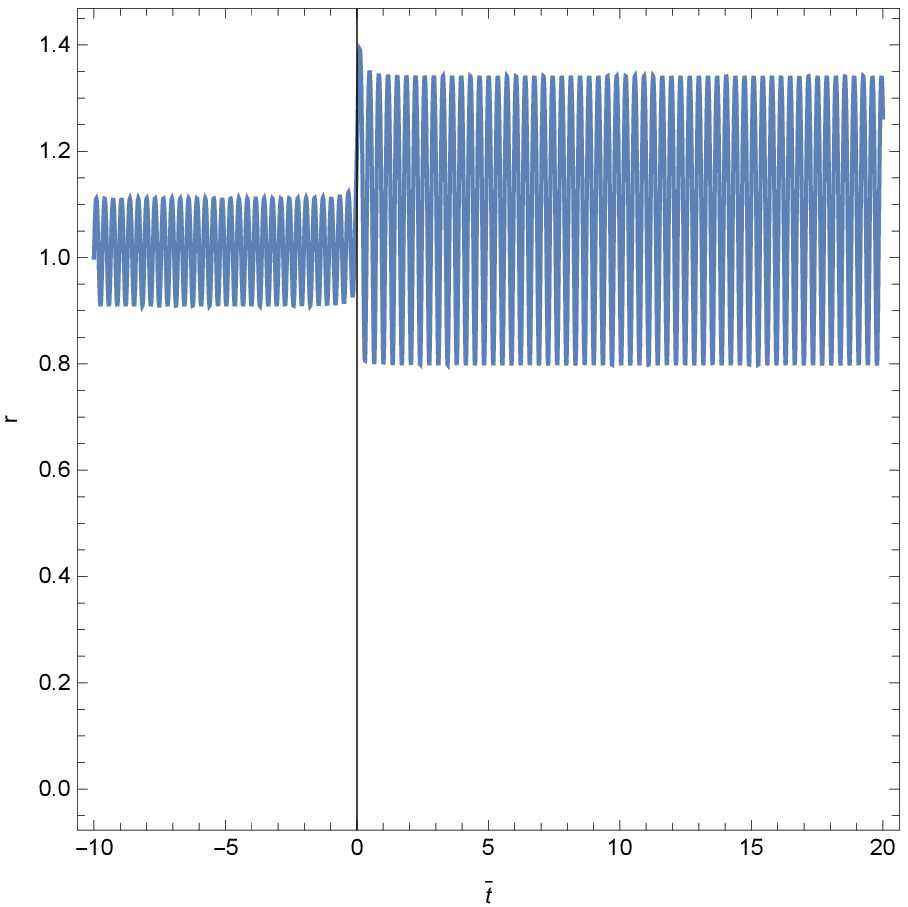}
\includegraphics[width=15pc]{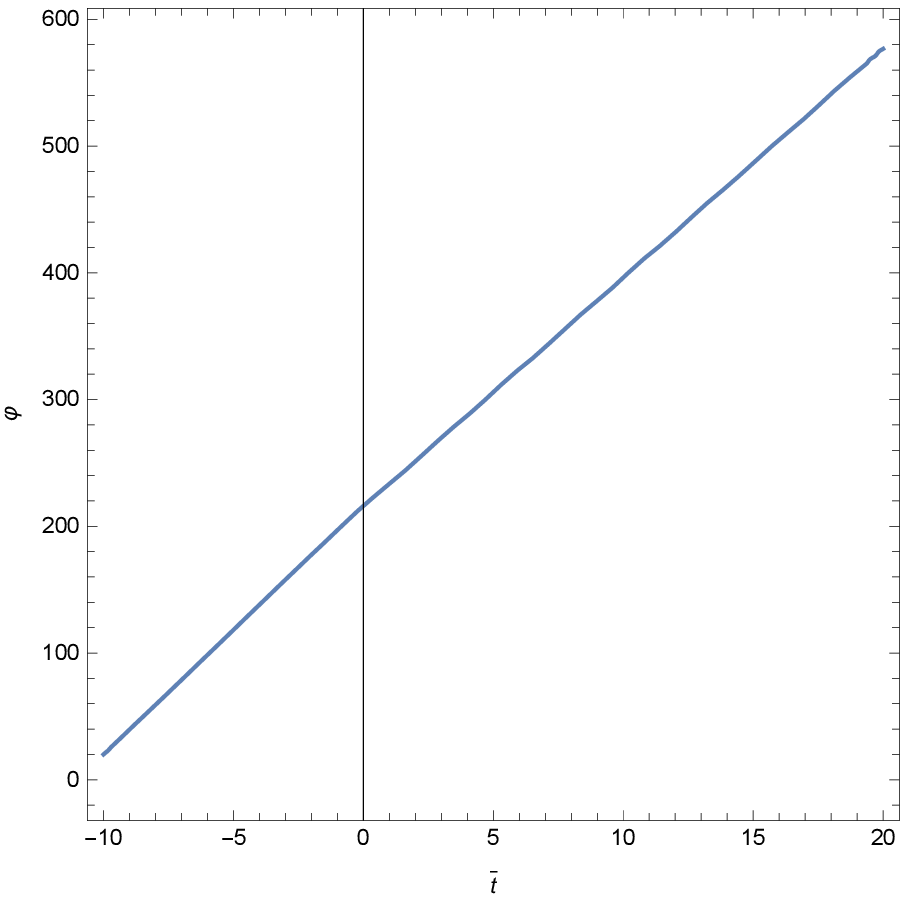}
\caption{Upper left plot: The trajectory of the point particle in
Cartesian coordinates around the gravitational source $M$, before
(blue curves), during and after (red curves) the Universe
experiences the pressure singularity. Upper right plot: The radial
coordinate $\bar{r}(t)$ as a function of the rescaled cosmic time.
It can clearly be seen that the characteristics of the elliptic
trajectory are affected by the pressure singularity at
$\bar{t}=0$}\label{plot1}
\end{figure}
Specifically, it would change the elliptic orbit of the moon
around the earth and this in effect would generate observable
effects on the surface of the earth. These effects would be the
same in origin to the effects of the tidal gravitational forces on
the surface of the earth. The most known effect of these tidal
forces is the tide in the sea water. Thus a pressure singularity
before $70-150\, $Myrs, may actually cause entire seas to
disappear from the surface of the earth, if the tidal forces
caused by the change in the moon's orbit were significant enough.
Intriguingly enough, the changes in Tethys Sea chronologically
occurred nearly $70-150\, $Myrs, so it is possible that the change
in the orbit of the moon around earth, or even the changes in the
trajectory of earth around the sun, gave rise to tidal and climate
phenomena before $70-150\, $Myrs. In order to have an idea how a
pressure singularity could have affected the trajectory of bound
objects, we shall use the paradigm of Ref.
\cite{Perivolaropoulos:2016nhp}, which was performed in the
context of simple GR. We also adopt the same GR approach in order
to make the argument independent from models of $F(R)$ gravity. In
Ref. \cite{Perivolaropoulos:2016nhp}, the initial trajectory was
considered to be a circle, we shall assume that the initial curve
is actually an ellipsis. Following Ref.
\cite{Perivolaropoulos:2016nhp}, see also \cite{Nesseris:2004uj},
and also adopting their notation, in a FRW Universe in the context
of GR, the energy density and pressure are $\rho(t)=\frac{3}{8\pi
G} \left( \frac{\dot a^2}{a^2}\right)$ and $p(t)=\frac{1}{8\pi G}
\left( 2\frac{\ddot a}{a} +\frac{\dot a^2}{a^2}\right)$,  thus the
pressure becomes singular for a sudden singularity. Assuming for
simplicity that the pressure singularity occurs at $t=0$ a scale
factor which can model a Type II singular evolution is the
following,
\begin{equation}
 a(t)=1+ c \vert t \vert^\eta\, . \label{scfactans}
\end{equation}
The pressure singularity occurs for $1<\eta <2$, and this is the
case we shall consider in this presentation too. Now, the metric
that describes the Newtonian limit of a spacetime near a
gravitational source $M$, in a flat FRW expanding Universe is
\cite{Perivolaropoulos:2016nhp,Nesseris:2004uj},
\begin{equation}
ds^2=\left(1-\frac{2GM}{a(t)\rho}\right)\cdot dt^2-a(t)^2\cdot
\left(d\rho^2+\rho^2\cdot (d\theta^2+sin^2\theta
d\varphi^2)\right)\, , \label{met}
\end{equation}
which is valid when $\frac{2GM}{a(t)\rho}\ll 1$. This condition
cannot be violated in pressure singularities, since the scale
factor is finite on these singularities. Replacing $ r=a(t) \cdot
\rho $, the geodesics equation for the metric (\ref{met}) are
\cite{Perivolaropoulos:2016nhp,Nesseris:2004uj},
\begin{equation}
(\ddot{r}-{\ddot{a}\over a}r)+{GM \over r^2}-r\dot{\varphi}^2=0,
\,\,\, r^2\dot{\varphi}=L\label{geodr}
 \end{equation}
for both the coordinates, and $L$ is the angular momentum per unit
mass, which is a conserved quantity. Combining equations
(\ref{geodr}), the radial equation of motion of the a point
particle around the gravitational source $M$ is
\cite{Perivolaropoulos:2016nhp,Nesseris:2004uj},
\begin{equation}
\ddot{r}={\ddot{a}\over a}r + {L^2 \over r^3}-{GM \over r^2}\, ,
\label{radeqm1}
\end{equation}
and in order to obtain dimensionless quantities, one needs to make
the rescalings \cite{Perivolaropoulos:2016nhp,Nesseris:2004uj}
${\bar r}\equiv {r\over {r_0}}$, ${\bar {\omega_0}}\equiv \omega_0
t_0$ and ${\bar t}\equiv {t\over t_0}$, where $r_0$ and $t_0$ are
arbitrary time and length scales. Upon the rescaling and also by
defining ${\dot \varphi}(t_0)= \omega_0\equiv {{GM}\over
{r_0^3}}$, the geodesics equation becomes
\cite{Perivolaropoulos:2016nhp,Nesseris:2004uj},
\begin{equation}
{\ddot {\bar r}}-{{\bar \omega_0}^2\over {{\bar r}^3}} + {{\bar
{\omega_0}}^2\over {{\bar r}^2}}-{{\ddot a}\over {a}}{\bar r}=0\,
. \label{dleqm}
\end{equation}
Hence, if the scale factor approaches a pressure singularity, it
will be of the form given in Eq. (\ref{scfactans}), so the
geodesics equation becomes,
\begin{equation}
{\ddot {\bar{r}}}={{ \bar{\omega}_0}^2\over {{ \bar{r}}^3}} - {{
{\bar{\omega_0}}}^2\over {{
\bar{r}}^2}}+\frac{c\;\eta(\eta-1)\;\vert
\bar{t}\vert^{\eta-2}}{(c\;\vert \bar{t}\vert^{\eta}+1)}\bar{r} \,
.\label{sfsgeod}
\end{equation}
We solved numerically the above differential equation for various
values of the parameter $\eta$ and $c$ focusing on values for
which the bound system is not destroyed, but is simply disrupted.
We chose the initial conditions in such a way so that the
trajectory of the test particle around the gravitational source
$M$ before the singularity was an ellipsis. The results of our
analysis are presented in Fig. \ref{plot1}. In the upper left plot
of Fig. \ref{plot1} we present the trajectory of the point
particle in Cartesian coordinates around the gravitational source
$M$, before, during and after the Universe experiences the
pressure singularity. As it can be seen in blue curves, the
trajectory of the point test particle before the singularity was
an ellipsis, and as the singularity is approached, the trajectory
is disrupted, starting at nearly $t=0$. At that point the red
curve describes the disruption of the trajectory of the point
particle around $M$, which after a while it also becomes an
ellipsis. In the upper right plot of Fig. \ref{plot1}, we plot the
radial coordinate as a function of the rescaled cosmic time. As it
can be seen, the elliptic trajectory is changed at the
singularity, and also the trajectory is an elliptic curve after
the singularity, with different characteristics though. In the
bottom of Fig. \ref{plot1} we plot the function $\varphi(t)$.
These simple considerations can be extended in the $F(R)$ gravity
case, for specific models of $F(R)$ gravity, but they suffice to
support our argument that pressure singularities might actually
have caused observable tidal or even climatological effects on
earth before $70-150\, $Myrs. These effects might changed the
surface of the earth, removing entire seas or even oceans from
their initial position, if the tidal effects of moon's gravity
were severe enough.

\section{Conclusions}

In this article we considered the scenario that an abrupt change
in physics occurred around $70-150\, $Myrs ago might have been
caused by a pressure singularity. The Universe can smoothly pass
through such a pressure singularity, since these are not crushing
type singularities, and the only physical quantity that diverges
is the effective pressure of the Universe everywhere on the three
dimensional spacelike hypersurface corresponding to the time
instance at which the singularity occurs. The strong energy
conditions are not violated in such types of singularities, and we
explained how such singularities might occur in the context of
$F(R)$ gravity. Also in $F(R)$ gravity, the strong energy
conditions can easily be satisfied. Moreover, we showed that a
pressure singularity can actually cause a blow-up in the $F(R)$
gravity effective gravitational constant, and this phenomenon is
also aligned with the abrupt physics changes argument of Refs.
\cite{Marra:2021fvf}. We should also note that there might be a
connection of our present approach with a maximal turn round
radius which is related to the shape of the specific $f(R)$
gravity, see for example \cite{Capozziello:2018oiw}. Apart from
this feature which may depend strongly on the specific $f(R)$
gravity model, there is a similarity between the pressure
singularity and the transition redshift that relates deceleration
to acceleration in cosmographic approaches
\cite{Capozziello:2014zda}. Certainly, these perspectives should
be closely discussed in some future article.

Coming back on the results of our article, as we discussed in the
text, a pressure singularity occurring before $70-150\, $Myrs,
could have disrupted the elliptic orbits of bound objects in the
Universe, without destroying them, but deforming these to ellipses
with different characteristics. As we showed numerically, this is
feasible in the presence of a pressure singularity. Such
deformation in the elliptic trajectories of bound objects could
also have occurred in our solar system, and could have had direct
effects on the surface of the earth or even in the climate of the
earth before $70-150\, $Myrs. Indeed, even deforming the orbit of
the moon could have had observable effects on oceans and sea
waters at that time. Also the same effects along with
climatological changes could have occurred due to the change of
the elliptic orbit of the earth around the sun. Thus earth's
primordial history might reveal such abrupt changes in the physics
of the Universe, as is also nicely pointed out in Ref.
\cite{Perivolaropoulos:2016nhp}. In the same line of research,
recently in Ref. \cite{Perivolaropoulos:2022vql} it was also
pointed out that the extinction of dinosaurs may be related to the
Hubble crisis. This argument further supports the general idea of
seeking imprints of the late-time Universe on the geological and
environmental history of earth. This line of research is
sensational, it might open up new horizons in theoretical
cosmology and astrophysics.

\section*{Acknowledgments}

We are grateful to Leandros Perivolaropoulos for his encouraging
comments and suggestions. This work was supported by MINECO
(Spain), project PID2019-104397GB-I00 (S.D.O).

\end{document}